\documentclass[prd,preprint,tightenlines,floatfix,showpacs,preprintnumbers,nofootinbib,eqsecnum]{revtex4}

 \usepackage[dvips,final]{graphicx}
  \usepackage{amssymb}
   \usepackage{amsmath}
    \usepackage{amsfonts}
     \usepackage{epsfig}
      \usepackage{bm}% bold math

\begin{document}

\title{Production of two $c \bar c$ pairs in double-parton scattering}

\author{Marta {\L}uszczak}
\email{luszczak@univ.rzeszow.pl} \affiliation{
University of Rzesz\'ow, PL-35-959 Rzesz\'ow, Poland}

\author{Rafa{\l} Maciu{\l}a}
\email{rafal.maciula@ifj.edu.pl} \affiliation{Institute of Nuclear
Physics PAN, PL-31-342 Cracow, Poland}

\author{Antoni Szczurek}
\email{antoni.szczurek@ifj.edu.pl} \affiliation{Institute of Nuclear
Physics PAN, PL-31-342 Cracow,
Poland and\\
University of Rzesz\'ow, PL-35-959 Rzesz\'ow, Poland}

\date{\today}

\begin{abstract}
We discuss production of two pairs of $c \bar c$ within a simple
formalism of double-parton scattering (DPS). Surprisingly very large cross sections,
comparable to single-parton scattering (SPS) contribution, are predicted for LHC
energies.
Both total inclusive cross section as a function of energy and
differential distributions for $\sqrt{s}$ are shown.
We discuss a perspective how to identify the double scattering
contribution.
\end{abstract}


\maketitle

%----------------------------
\section{Introduction}
%----------------------------

It is commonly believed that gluon-gluon fusion is the dominant
mechanism of $c$ or $\bar c$ production at high energies.
Then in leading-order (LO) approximation the differential cross section for the single-parton scattering (SPS)
production of heavy quark and heavy antiquark pair reads:
\begin{equation}
\frac{d \sigma}{d y_1 d y_2 d^2p_t} = \frac{1}{16 \pi^2 {\hat s}^2}
x_1 g(x_1,\mu^2) \; x_2 g(x_2,\mu^2) \;
\overline{|{\cal M}_{gg \to Q \bar Q}|^2} \; ,
\label{LO_collinear_gg}
\end{equation}
where longitudinal momentum fractions can be calculated from kinematical
variables of final quark and antiquark as: 
$x_1 = \frac{m_t}{\sqrt{s}}(\exp( y_1) + \exp( y_2))$,
$x_2 = \frac{m_t}{\sqrt{s}}(\exp(-y_1) + \exp(-y_2))$ 
with $y$'s being quark (antiquark) rapidities and $m_t$ being a quark (antiquark) transverse mass.
We have limited here to gluon-gluon fusion only which is the dominant mechanism
at high energies. The quark-antiquark annihilation plays some role
only close to the kinematical threshold and/or large rapidities. In general, the higher-order
corrections do not change most of observables leading to a 
rough renormalization of the cross section by the so-called $K$ factor.

\begin{figure}[!h]
\begin{minipage}{0.4\textwidth}
 \centerline{\includegraphics[width=1.0\textwidth]{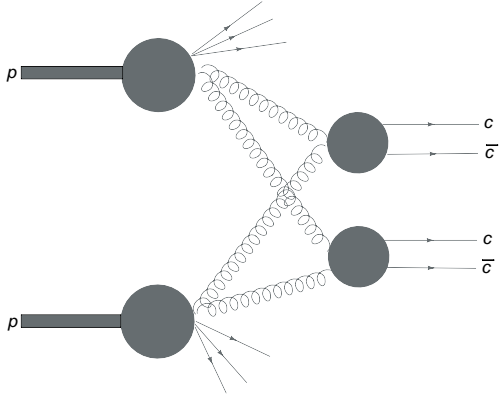}}
\end{minipage}
   \caption{
\small Mechanism of $c \bar c c \bar c$ production via double-parton 
scattering. 
}
 \label{fig:diagram}
\end{figure}

In the present paper we wish to estimate the contribution of 
double-parton scatterings (DPS).
The mechanism of double-parton scattering production of two pairs of
heavy quark and heavy antiquark is shown in Fig.~\ref{fig:diagram}.

The double-parton scattering has been recognized and discussed already 
in seventies and eighties 
\cite{LP78,T1979,GSH80,H1983,PT1984,PT1985,M1985,HO1985,SZ1987}. 
The activity stopped when it was realized that their contribution at those times
available center-of-mass energies was negligible.
Several estimates of the cross section for different processes have been
presented in recent years \cite{DH1996,KS2000,FT2002,BJS2009,GKKS2010,
SV2011,BDFS2011,KKS2011,BSZ2011}. The theory of the double-parton scattering is quickly developing
(see e.g. \cite{S2003,KS2004,SS2004,GS2010,GS2011,DS2011,RS2011,DOS2011}).

In the present analysis we wish to concentrate on the production of $(c \bar c)
(c \bar c)$ four-parton final state which has not been carefully discussed so
far, but, as will be shown here, is particularly interesting especially in 
the context of experiments being carried out at LHC and/or high-energy atmospheric
and cosmogenic neutrinos (antineutrinos).

The double-parton scattering formalism proposed so far assumes two
single-parton scatterings. Then in a simple probabilistic picture the cross section for double-parton 
scattering can be written as:
\begin{equation}
\sigma^{DPS}(p p \to c \bar c c \bar c X) = \frac{1}{2 \sigma_{eff}}
\sigma^{SPS}(p p \to c \bar c X_1) \cdot \sigma^{SPS}(p p \to c \bar c X_2).
\label{basic_formula}
\end{equation}
This formula assumes that the two subprocesses are not correlated and do not interfere.
At low energies one has to include parton momentum conservation
i.e. extra limitations: $x_1+x_3 <$ 1 and $x_2+x_4 <$ 1, where $x_1$ and $x_3$
are longitudinal momentum fractions of gluons emitted from one proton and $x_2$ and $x_4$
their counterpairs for gluons emitted from the second proton. The "second"
emission must take into account that some momentum was used up in the first parton
collision. This effect is important at large quark or antiquark rapidities.
Experimental data \cite{Tevatron} provide an estimate of $\sigma_{eff}$
in the denominator of formula (\ref{basic_formula}). In our analysis we
take a rather conservative value $\sigma_{eff}$ = 15 mb.

%-----------------------------------------------------------------------------
%------------------------------------------------------------------------------

The simple formula (\ref{basic_formula}) can be generalized to include 
differential distributions. Again in leading-order approximation 
differential distribution can be written as
\begin{equation}
\frac{d \sigma}{d y_1 d y_2 d^2 p_{1t} d y_3 d y_4 d^2 p_{2t}}  \\ =
\frac{1}{ 2 \sigma_{eff} }
\frac{ d \sigma } {d y_1 d y_2 d^2 p_{1t}} \cdot
\frac{ d \sigma } {d y_3 d y_4 d^2 p_{2t}} 
\label{differential_distribution}
\end{equation}
which by construction reproduces formula for integrated cross section (\ref{basic_formula}).
This cross section is formally differential in 8 dimensions but can be 
easily reduced to 7 dimensions noting that physics of unpolarized
scattering cannot depend on azimuthal angle of the pair or on azimuthal angle of one of 
the produced $c$ ($\bar c$) quark (antiquark).
The differential distributions for each single scattering step can be written in terms
of collinear gluon distributions with longitudinal momentum fractions
$x_1$, $x_2$, $x_3$ and $x_4$ expressed in terms of rapidities $y_1$, $y_2$, $y_3$,
$y_4$ and transverse momenta of quark (or antiquark) for each step 
(in the LO approximation identical for quark and antiquark).

A more general formula for the cross section can be written formally 
in terms of double-parton distributions, e.g. $F_{gg}$, $F_{qq}$, etc. 
In the case of heavy quark (antiquark) production at high energies:
\begin{eqnarray}
d \sigma^{DPS} &=& \frac{1}{2 \sigma_{eff}}
F_{gg}(x_1,x_2,\mu_1^2,\mu_2^2) F_{gg}(x'_{1}x'_{2},\mu_1^2,\mu_2^2)
\nonumber \\
&&d \sigma_{gg \to c \bar c}(x_1,x'_{1},\mu_1^2)
d \sigma_{gg \to c \bar c}(x_2,x'_{2},\mu_2^2) \; dx_1 dx_2 dx'_1 dx'_2 \, .
\label{cs_via_doublePDFs}
\end{eqnarray}
It is physically motivated to write the double-parton distributions rather in the impact
parameter space $F_{gg}(x_1,x_2,b) = g(x_1) g(x_2) F(b)$, where $g$ are 
usual conventional parton distributions and $F(b)$ is an overlap of the matter
distribution in the 
transverse plane where $b$ is a distance between both gluons in the
transverse plane \cite{CT1999}. The effective cross section in 
(\ref{basic_formula}) is then $1/\sigma_{eff} = \int d^2b F^2(b)$ and 
in this approximation is energy independent.

The double-parton distributions in Eq.(\ref{cs_via_doublePDFs})
are generally unknown. Usually one assumes a factorized form and
expresses them via standard distributions for SPS.
Even if factorization is valid at some scale, QCD evolution may lead
to a factorization breaking. Evolution is known only in the case 
when the scale of both scatterings is the same \cite{S2003,KS2004,GS2010}
i.e. for heavy object, like double gauge boson production.
For double $c \bar c$ production this is not the case and was
not discussed so far in the literature. In the present preliminary study
we shall therefore apply the commonly used in the literature
factorized model. A refinement will be done elsewhere. In explicit calculations
presented below we use leading order collinear gluon distributions
(GRV94 \cite{GRV94}, CTEQ6 \cite{CTEQ6}, GJR08 \cite{GJR08}, MSTW08 \cite{MSTW08}).

%--------------------
\section{Results}
%--------------------

\begin{figure}[!h]
\begin{minipage}{0.47\textwidth}
 \centerline{\includegraphics[width=1.0\textwidth]{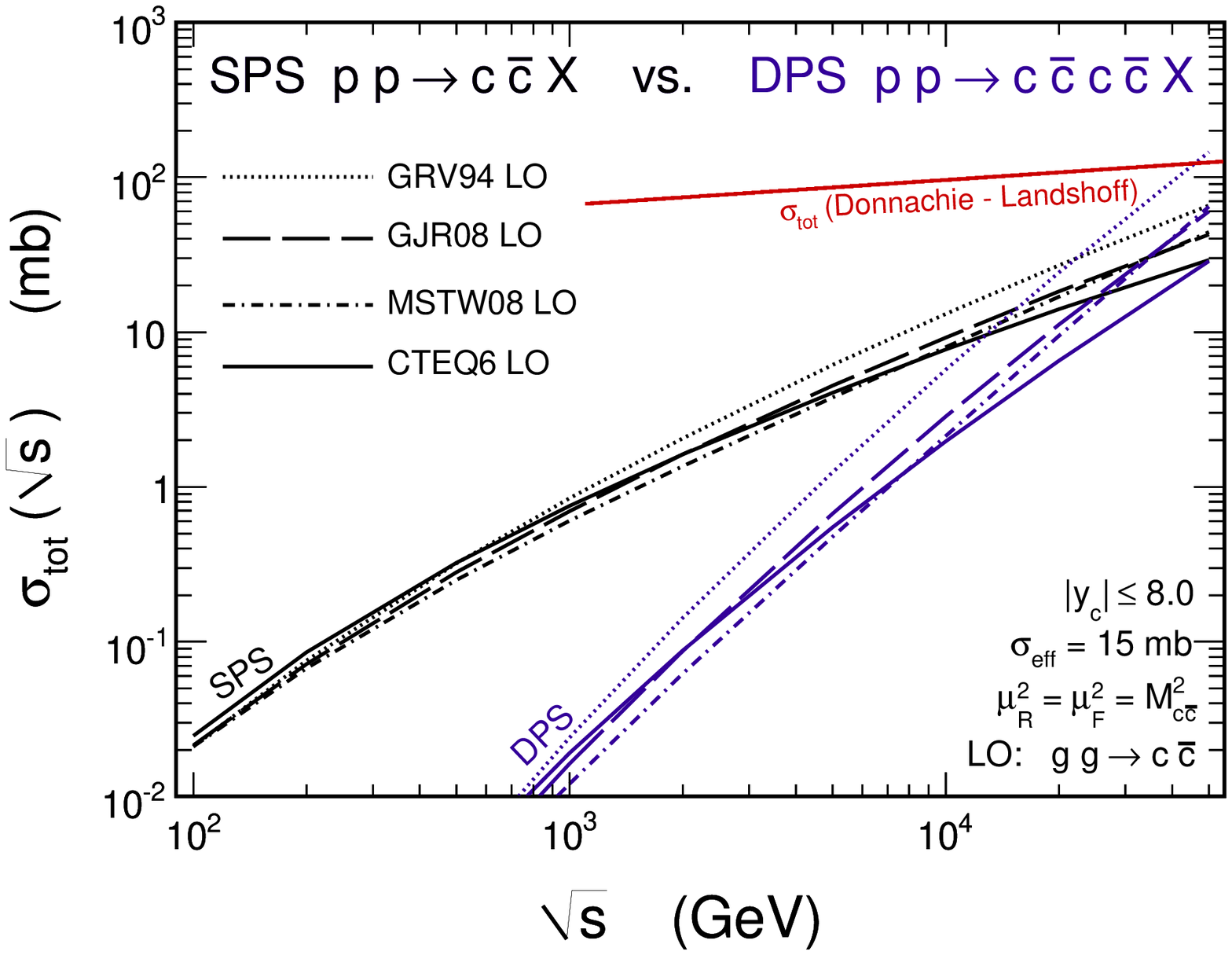}}
\end{minipage}
\hspace{0.5cm}
\begin{minipage}{0.47\textwidth}
 \centerline{\includegraphics[width=1.0\textwidth]{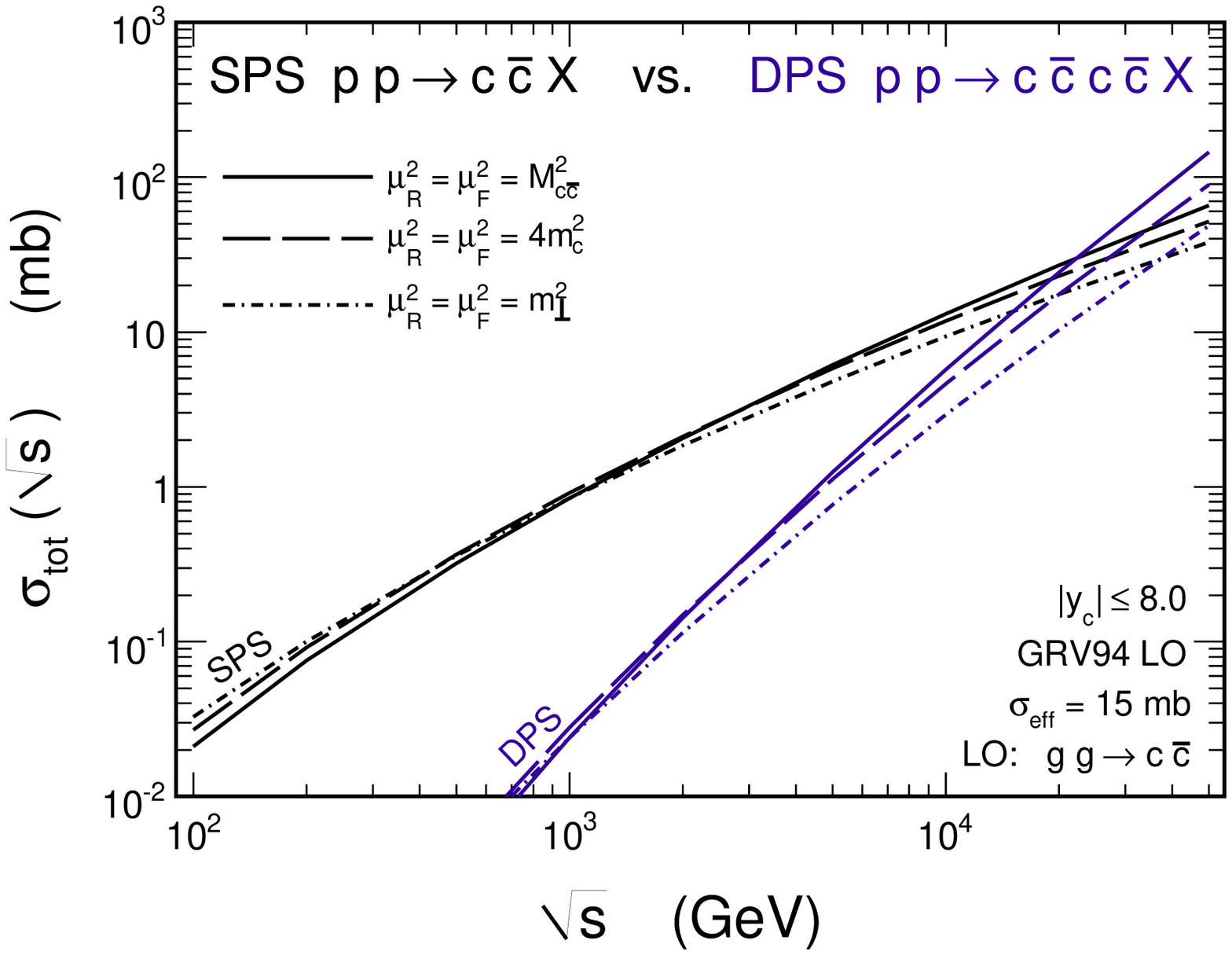}}
\end{minipage}
   \caption{
\small Total LO cross section for single-parton and double-parton
scattering as a function of center-of-mass energy (left panel) and 
uncertainties due to the choice of (factorization, renormalization) scales (right panel). 
We show in addition a parametrization of the total cross section in the left panel.
Cross section for DPS should be multiplied in addition by a factor 2 in the case when all $c$ ($\bar c$) are counted. 
}
 \label{fig:single_vs_double_LO}
\end{figure}

In Fig.~\ref{fig:single_vs_double_LO} we
compare cross sections for the single and double-parton
scattering as a function of proton-proton center-of-mass energy. At low energies the 
conventional single-parton scattering dominates. For reference we show the proton-proton total cross section as a function
of energy as parametrizes in Ref.~\cite{DL92}.
At low energy the $c \bar c$ or $ c \bar c c \bar c$ cross sections are much
smaller than the total cross section. At higher energies the contributions
dangerously approach the expected total cross section\footnote{New experiments at LHC will provide new input for parametrizations of the total cross section.}. This shows that inclusion of
unitarity effect and/or saturation of parton distributions may be necessary.
The effect of saturation in $c\bar c$ production has been included e.g. in Ref.~\cite{enberg1}
but not checked versus experimental data. Presence of double-parton scattering changes the situation. The double-parton scattering
is therefore potentially very important ingredient in the context of high energy neutrino production
in the atmosphere \cite{GIT96, MRS2003, enberg1} or of cosmogenic origin \cite{enberg2}.
We leave this rather difficult issue for future studies where the LHC charm data must be included.
 At LHC energies the
cross section for both terms become comparable\footnote{If inclusive cross
section for $c$ or $\bar c$ was shown the cross section should be
multiplied by a factor of two -- two $c$ or two $\bar c$ in each event.}.
This is a completely new situation
when the double-parton scattering gives a huge contribution to inclusive
charm production. 

%-----------------------------------------------------------------------------
%-----------------------------------------------------------------------------

\begin{figure}[!h]
\begin{minipage}{0.47\textwidth}
 \centerline{\includegraphics[width=1.0\textwidth]{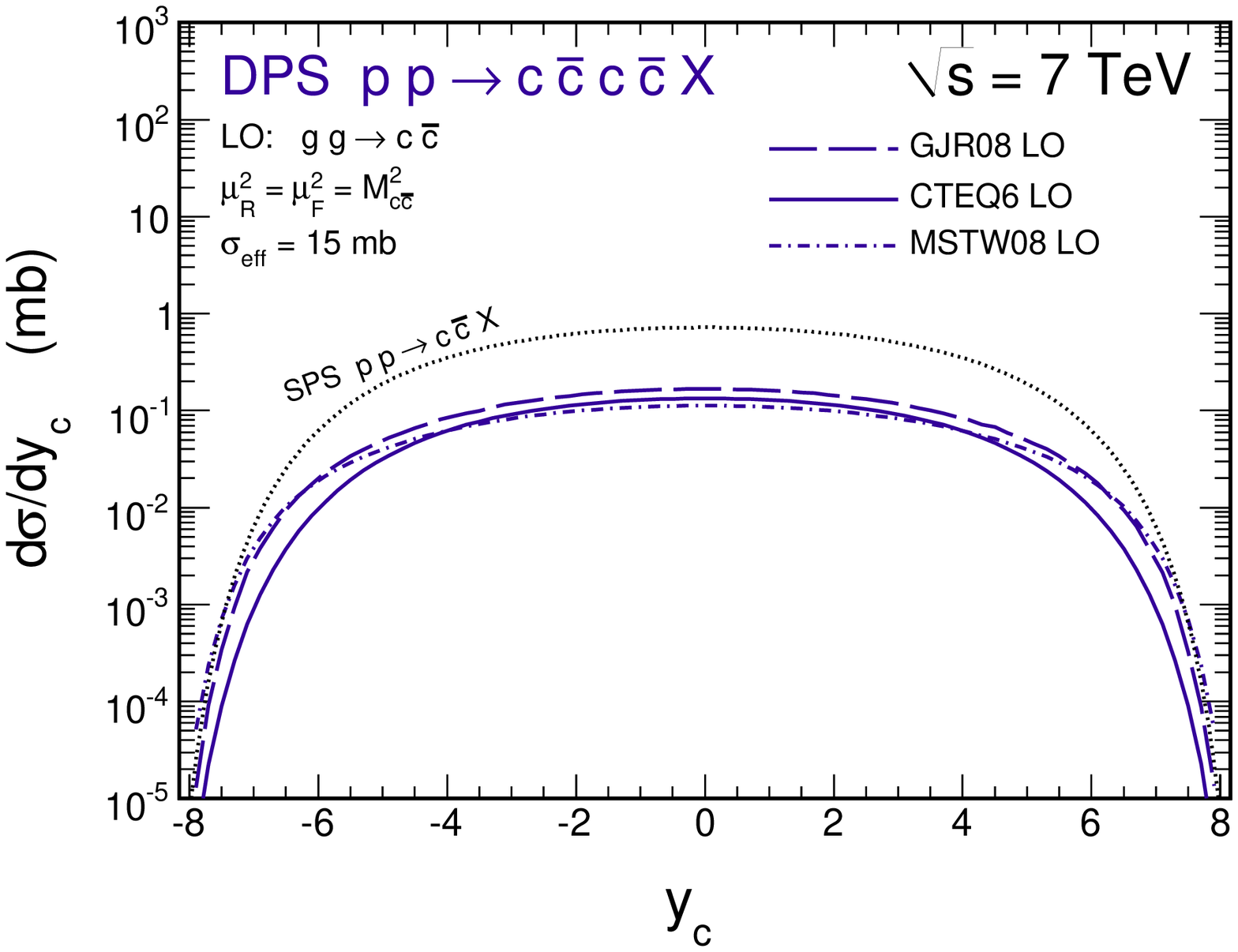}}
\end{minipage}
\hspace{0.5cm}
\begin{minipage}{0.47\textwidth}
 \centerline{\includegraphics[width=1.0\textwidth]{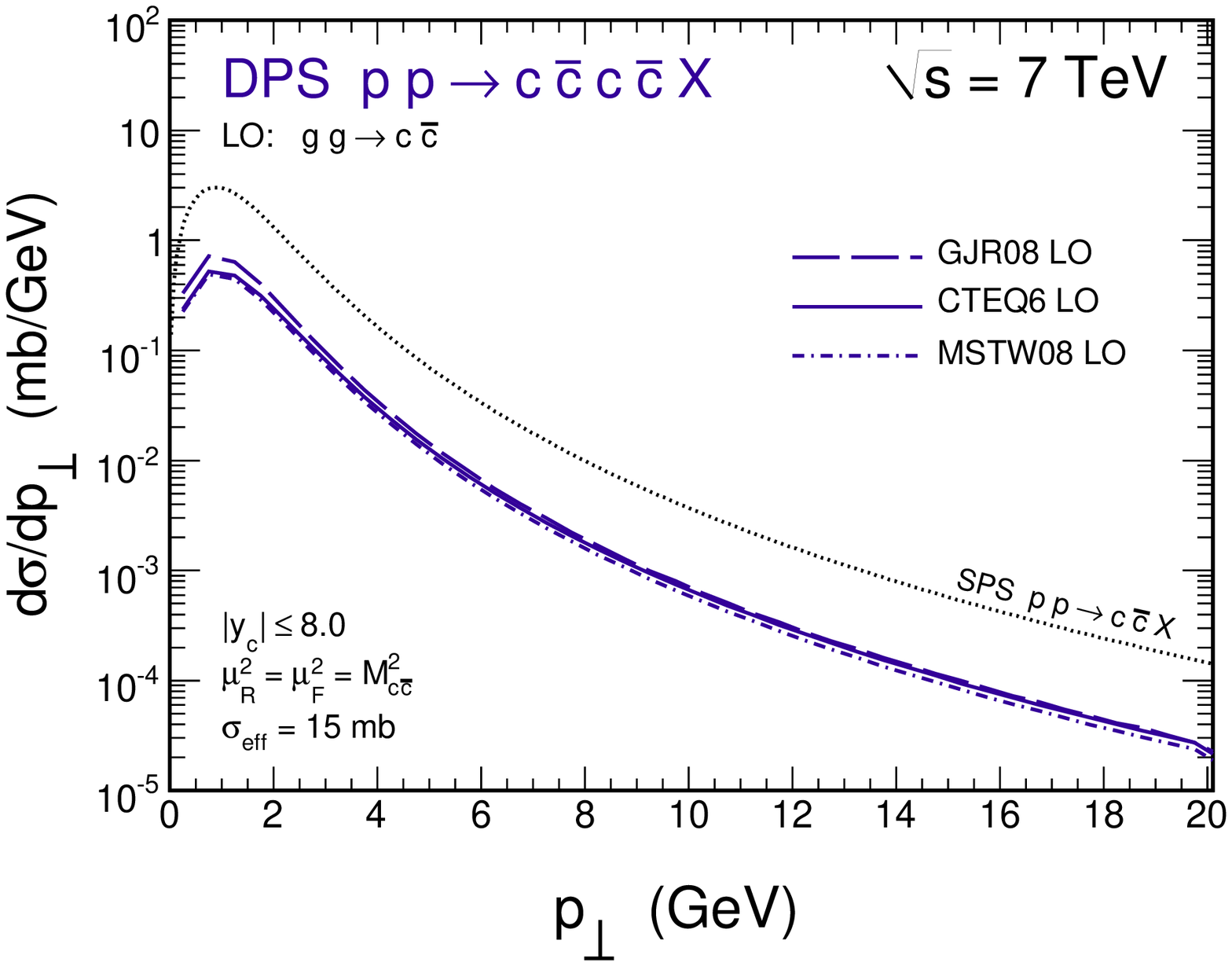}}
\end{minipage}
   \caption{
\small Distribution in rapidity (left panel) and transverse momentum (right panel) of 
$c$ or $\bar{c}$ quarks  at $\sqrt{s}$ = 7 TeV. Cross section for DPS should be multiplied in addition by a factor 2 in the case when all $c$ ($\bar c$) are counted. 
}
 \label{fig:double_single1}
\end{figure}

\begin{figure}[!h]
\begin{minipage}{0.47\textwidth}
 \centerline{\includegraphics[width=1.0\textwidth]{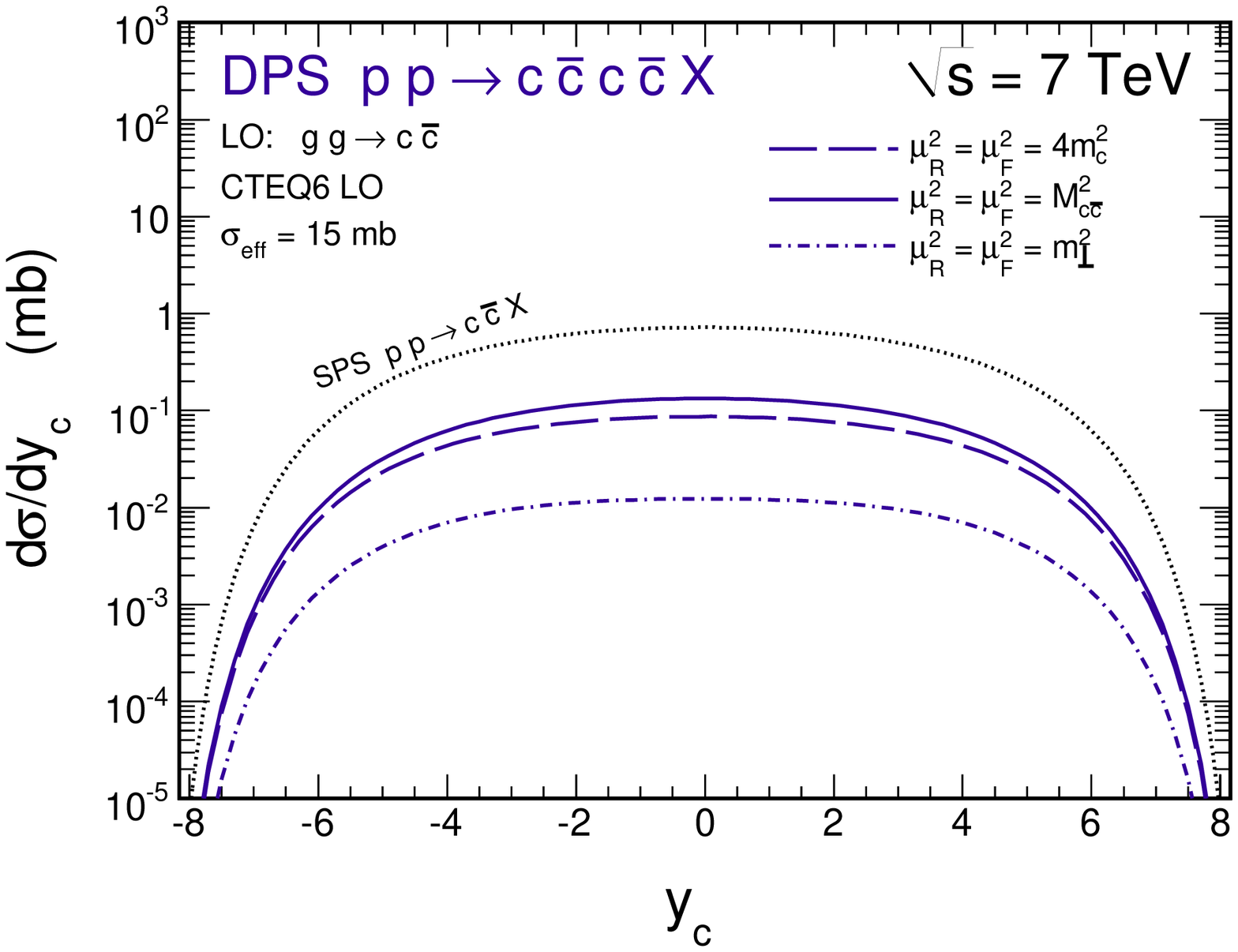}}
\end{minipage}
\hspace{0.5cm}
\begin{minipage}{0.47\textwidth}
 \centerline{\includegraphics[width=1.0\textwidth]{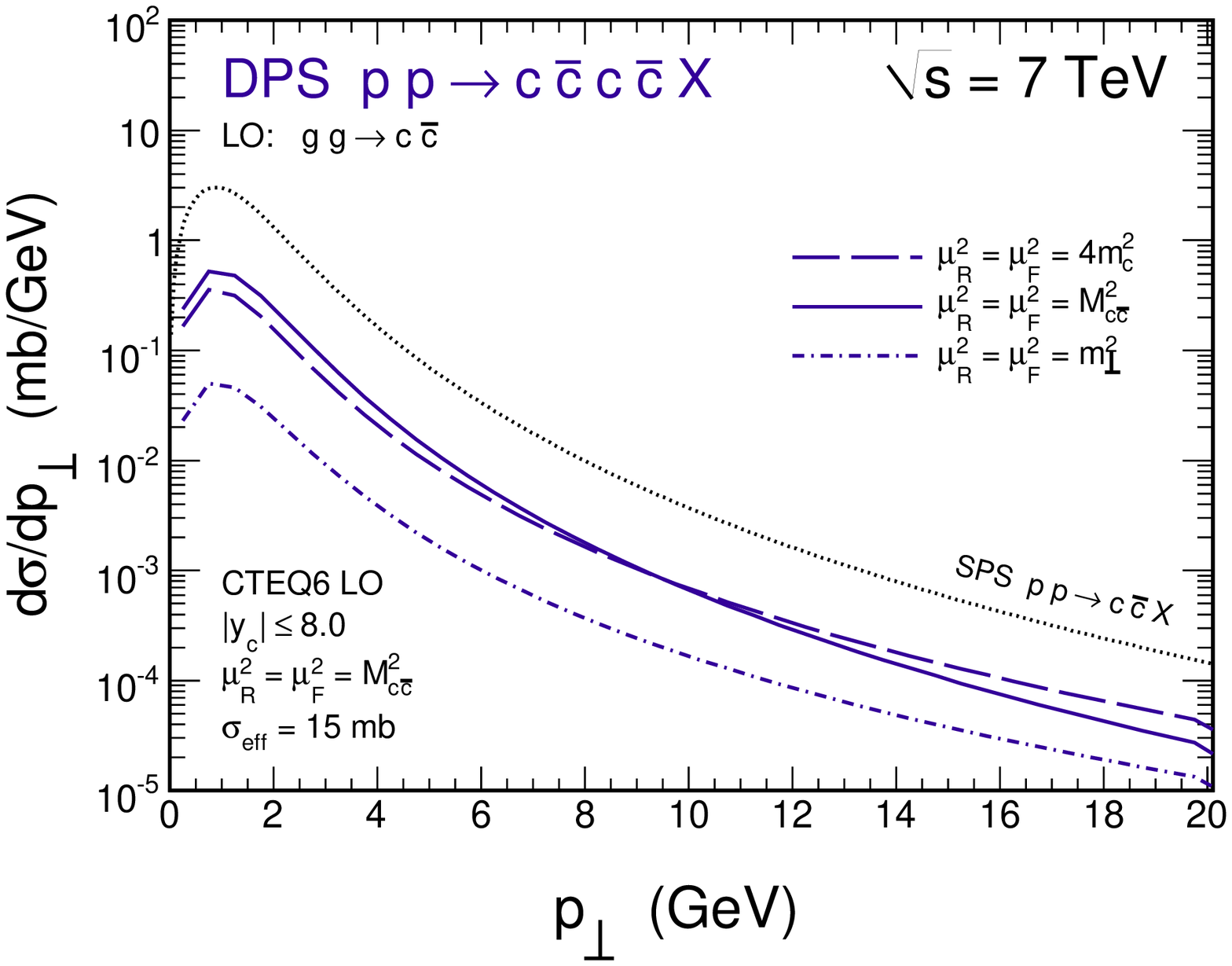}}
\end{minipage}
   \caption{
\small Uncertainties related to renormalization and factorization
scales choice for distributions in rapidity (left panel) and transverse 
momentum (right panel) of $c$ or $\bar{c}$ quarks  at $\sqrt{s}$ = 7 TeV. Cross section for DPS should be multiplied in addition by a factor 2 in the case when all $c$ ($\bar c$) are counted. 
}
 \label{fig:double_single2}
\end{figure}

In Figs.~\ref{fig:double_single1}, \ref{fig:double_single2},  we present
single $c$ ($\bar c$) distributions. Within approximations made in this
paper the distributions are identical in shape to single-parton
scattering distributions. This means that double-scattering contribution
produces naturally an extra center-of-mass energy dependent $K$ factor
to be contrasted with approximately energy-independent $K$-factor due to
next-to-leading order corrections. One can see a strong dependence on 
the factorization and renormalization scales which produces almost 
order-of-magnitude uncertainties and precludes a more precise estimation. 
A better estimate could be done when LHC charm data are published  and 
the theoretical distributions are somewhat adjusted to experimental data.

%-----------------------------------------------------------------------------
%-----------------------------------------------------------------------------

%-----------------------------------------------------------------------------
%-----------------------------------------------------------------------------

\begin{figure}[!h]
\begin{minipage}{0.47\textwidth}
 \centerline{\includegraphics[width=1.0\textwidth]{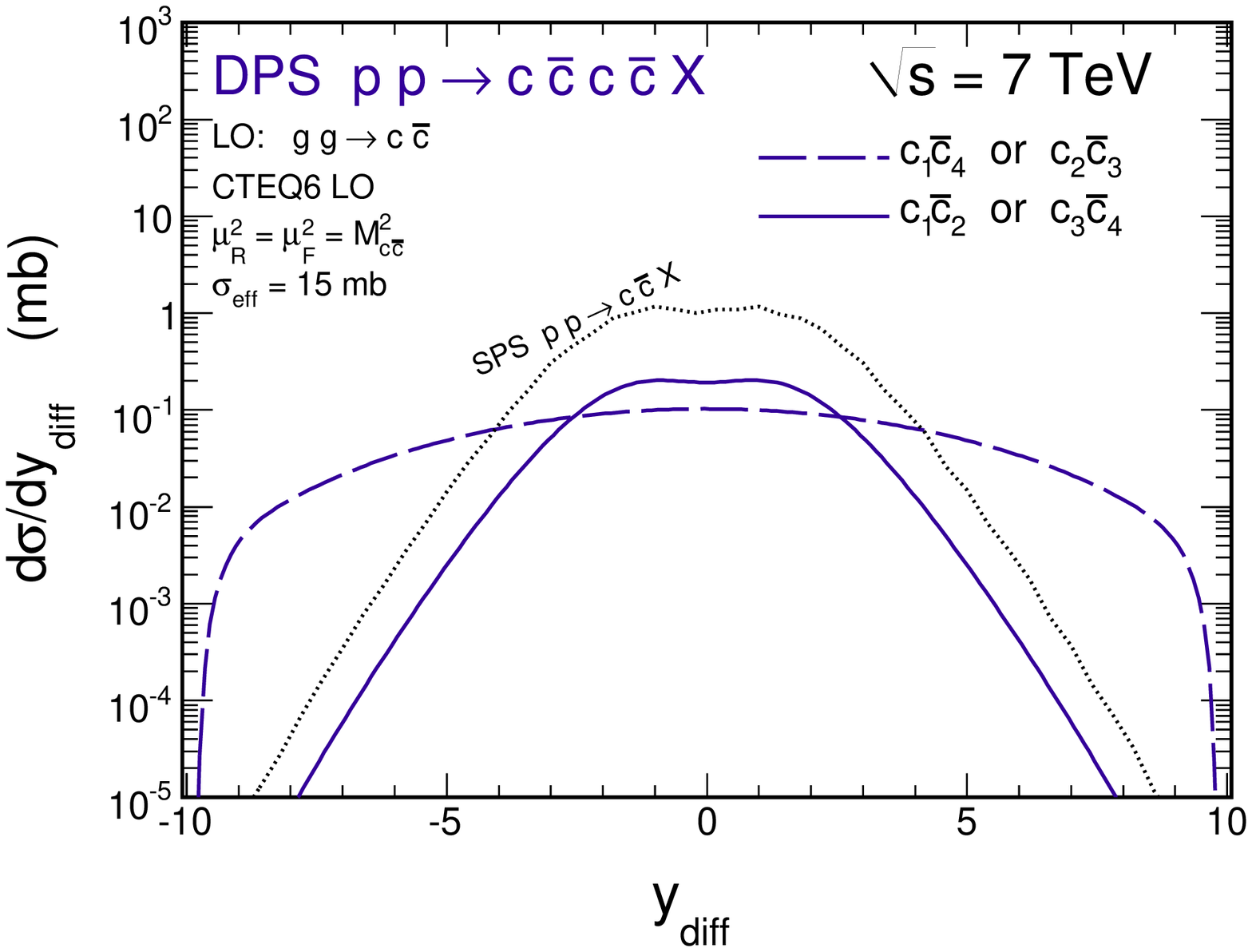}}
\end{minipage}
\hspace{0.5cm}
\begin{minipage}{0.47\textwidth}
 \centerline{\includegraphics[width=1.0\textwidth]{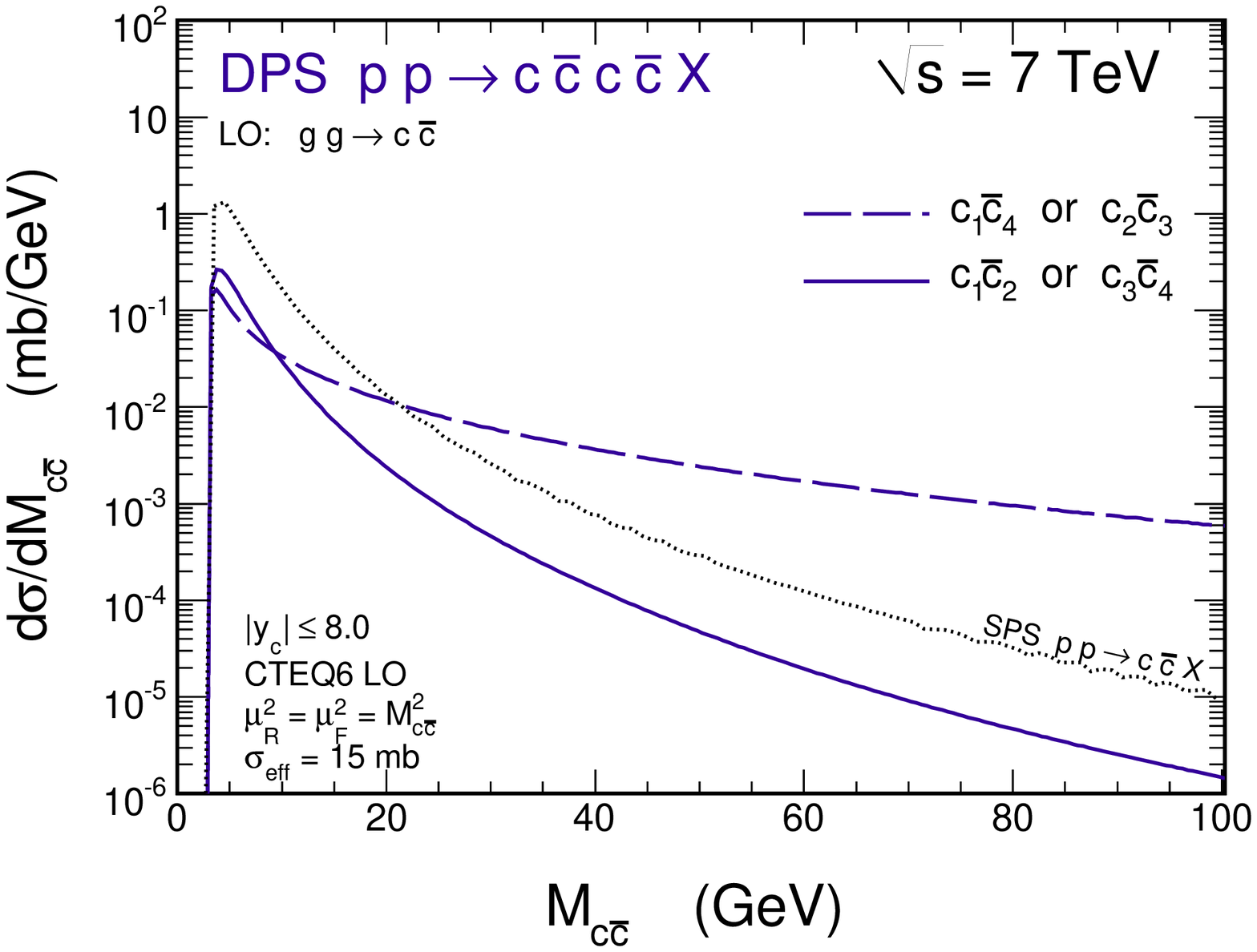}}
\end{minipage}
   \caption{
\small Distribution in rapidity difference (left panel) and in invariant
mass of the $c\bar{c}$ pair (right panel) at $\sqrt{s}$ = 7 TeV.
}
\label{fig:double_correlations_1}
\end{figure}

So far we have discussed only single particle spectra of $c$ or $\bar c$ 
(rapidity, transverse momentum distributions) which due to scale
dependence do not provide a clear test of the existence of double-parton
scattering contributions.
A more stringent test could be performed by studying correlation
observables.
In particular, correlations between $c$ and $\bar c$ are very interesting
even without double-parton scattering terms \cite{LS2006}.
In Fig.~\ref{fig:double_correlations_1} we show distribution in the
difference of $c$ and $\bar c$ rapidities $y_{diff} = y_c - y_{\bar c}$ (left panel) as well as in the $c \bar c$
invariant mass $M_{c\bar c}$ (right panel). We show both terms: when $c \bar c$ are emitted
in the same parton scattering ($c_1\bar c_2$ or $c_3\bar c_4$) and when they are emitted from different 
parton scatterings ($c_1\bar c_4$ or $c_2\bar c_3$). In the latter case we observe a long tail for large
rapidity difference as well as at large invariant masses of $c \bar c$.
Such distributions cannot be directly measured for $c \bar c$ but could
be measured for mesons (rapidity difference up to 5 for the main ATLAS or CMS
detector) or electron-positron or $\mu^+ \mu^-$. The ALICE forward muon
spectrometer \cite{ALICE} covers the pseudorapidity interval 2.5 $< \eta <$ 4
which when combined with the central detector means pseudorapidities 
differences up to 5. This is expected to be a region of phase space where
double-parton scattering contribution would most probably dominate over 
single-parton scattering contribution. This will be a topic of a 
forthcoming analysis.
Next-to-leading order corrections are not expected to give
major contribution at large pseudorapidity differences or/and
large invariant masses of $\mu^+ \mu^-$ but this must be verified in the future.
The CMS detector is devoted especially to measurements of muons. The lower
transverse momentum threshold is however rather high, the smallest being
about 1.5 GeV at $\eta = \pm$ (2 - 2.4) which may be interesting for
double-parton scattering searches. This requires special dedicated Monte Carlo
studies.

%-----------------------------------------------------------------------------
%------------------------------------------------------------------------------

\begin{figure}[!h]
\begin{minipage}{0.47\textwidth}
 \centerline{\includegraphics[width=1.0\textwidth]{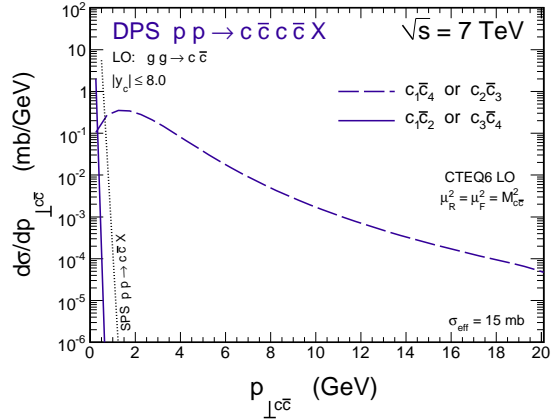}}
\end{minipage}
   \caption{
\small Distribution in transverse momentum of $c\bar{c}$ pairs from the
same parton scattering and from different parton scatterings at 
$\sqrt{s}$ = 7 TeV. 
}
 \label{fig:double_correlations_2}
\end{figure}

Finally in Fig.~\ref{fig:double_correlations_2} we present distribution
in the transverse momentum of the $c \bar c$ pair $|\overrightarrow{p_{\perp c\bar c}}|$, where $\overrightarrow{p_{\perp c\bar c}} = \overrightarrow{p_{\perp c}} + \overrightarrow{p_{\perp \bar c}}$ which is a Dirac delta function
in the leading-order approximation. In contrast, double-parton
scattering mechanism provide a broad distribution extending to large 
transverse momenta. Next-to-leading order corrections obviously
destroy the $\delta$-like leading-order correlation. We believe that
similar distributions for $D \bar D$ or/and $e^+ e^-$ or $\mu^+ \mu^-$ pairs
would be a useful observables to identify the DPS contributions
but this requires real Monte Carlo simultions including actual limitations of
experimental apparatus. Correlations between outgoing nonphotonic
electrons has been studied at much lower RHIC energy in Ref.~\cite{MSS2011}.
 
%-----------------------------------------------------------------------------
%------------------------------------------------------------------------------

Production of two $c \bar c$ pairs in the leading order approximation
is only a first step in trying to identify DPS contribution.
In the next step we plan next-to-leading order calculation of the same process.
Inclusion of hadronization and/or semileptonic decays would be very useful in planning experimental searches.

\vspace{1cm}

{\bf Acknowledgments}

We are indebted to Wolfgang Sch\"afer for interesting
conversion and Jacek Oko{\l}owicz for careful reading of this manuscript. This work was partially supported by the polish
grant N N202 237040.

%--------------------------------------------------------------------

\end{document}